\documentclass[12pt,preprint]{aastex}

\newcommand{\kms}{km~s$^{-1}$}

\newcommand{\etal}{{\it et al.}}
\newcommand{\eg}{{\it e.g., }}

\shorttitle{SN 2001gd}

\slugcomment{Accepted to the Astrophysical Journal}

\shortauthors{Stockdale, \etal}

\begin{document}

\title{The Radio Evolution of SN 2001gd} 

\author{Christopher J. Stockdale}
\affil{Marquette University, Physics Department, PO Box 1881, Milwaukee, WI, 53214-1881, christopher.stockdale@mu.edu}

\author{Christopher L.~Williams\altaffilmark{1}}
\affil{Naval Research Laboratory, Code 7213, Washington, DC 20375-5351, clmw@mit.edu}

\author{Kurt W. Weiler}
\affil{Naval Research Laboratory, Code 7210, Washington, DC 20375-5320, kurt.weiler@nrl.navy.mil}

\author{Nino Panagia\altaffilmark{2,}\altaffilmark{3}}
\affil{Space Telescope Science Institute, 3700 San Martin Drive, Baltimore, MD 
21218, panagia@stsci.edu}

\author{Richard A. Sramek}
\affil{National Radio Astronomy Observatory, P.O.~Box 0, Socorro, NM 87801, dsramek@nrao.edu}

\author{Schuyler D. Van Dyk}
\affil{IPAC/Caltech, Mail Code 100-22, Pasadena, CA 91125, vandyk@ipac.caltech.edu}

\and

\author{Matthew T. Kelley}
\affil{Marquette University, Physics Department, PO Box 1881, Milwaukee, WI, 53214-1881, matthew.kelley@mu.edu}

\altaffiltext{1}{Present address: Massachusetts Institute of Technology,
Kavli Institute for Astrophysics and Space Research, Cambridge, MA
02139}
\altaffiltext{2}{INAF-Osservatorio Astrofisico di Catania, Via S. Sofia 78, I-95128 Catania, Italy}
\altaffiltext{3}{Supernova Ltd., Olde Yard Village \#131, Northsound
Road, Virgin Gorda, British Virgin Islands.}

\begin{abstract}
We present the results of observations of the radio emission from
Supernova 2001gd in NGC5033 from 2002 February 8 through 2006 September
25.  The data were obtained using the Very Large Array at wavelengths of
1.3 cm (22.4 GHz), 2 cm (14.9 GHz), 3.6 cm (8.4 GHz), 6 cm (4.9 GHz),
and 20 cm (1.5 GHz), with one upper limit at 90 cm (0.3 GHz). In
addition, one detection has been provided by the Giant Metrewave Radio
Telescope at 21 cm (1.4 GHz).  SN~2001gd was discovered in the optical
well past maximum light, so that it was not possible to obtain many of
the early radio ``turn-on" measurements which are important for
estimating the local circumstellar medium (CSM) properties. Only at 20
cm were turn-on data available.  However, our analysis and fitting of
the radio light curves, and the assumption that the Type IIb SN~2001gd
resembles the much better studied Type IIb SN~1993J, enables us to
describe the radio evolution as being very regular through day $\sim$550
and consistent with a nonthermal-emitting model with a thermal absorbing
CSM. The presence of synchrotron-self absorption (SSA) at early times is
implied by the data, but determination of the exact relationship between
the SSA component from the emitting region and the free-free absorption
component from the CSM is not possible as there are insufficient early
measurements to distinguish between models. After day $\sim550$, the
radio emission exhibits a dramatically steeper decline rate which,
assuming similarity to SN~1993J, can be described as an exponential
decrease with an e-folding time of 500 days. We interpret this abrupt
change in the radio flux density decline rate as implying a transition
of the shock front into a more tenuous region of circumstellar material.
A similar change in radio evolution has been seen earlier in other SNe
such as SN~1988Z, SN~1980K, and SN~1993J.
\end{abstract}

\keywords{supernovae, individual (SN 2001gd), galaxies, individual (NGC5033), radio continuum, 
stellar evolution: massive stars}

\section{Introduction}

Supernova (SN) 2001gd was discovered at magnitude 16.5 on 2001 November 24.820 UT by \citet{Nakano01}, located $\sim3'$ north-northwest of the center of NGC~5033.  \citet{Nakano01} report a position of $\alpha = 13^h 13^m 23\fs89$, $\delta = +36\arcdeg 38\arcmin 17\farcs7$ (J2000), for the SN and note that nothing was detected at this position in previous CCD images taken 179 days earlier (2001 May 29) to a limiting magnitude of 19.  Ten days later, \citet{Matheson01} obtained a spectrum of the SN, identifying it as a Type IIb SN well past its maximum light, with weak helium lines and nebular-phase lines of magnesium, oxygen, and calcium beginning to dominate.  They noted that this spectrum was nearly identical to that of Type IIb SN~1993J, on day 93 after its explosion.  Using this suggestion to constrain the explosion date for the SN, we assume that the optical discovery occured $\sim83$ days after the explosion and, therefore, for calculating the age of the supernova we assume an explosion date of 2001 September 03, which is the same explosion date assumed by \citet{PerezTorres05}.

\citet{Tully88} gives the distance to NGC~5033, as $18.7$ Mpc (calculated with $H_0 = 75~{\rm km}~{\rm s}^{-1}~{\rm Mpc}^{-1}$). Accepting their distance calculation and correcting it with an updated value of the Hubble constant $H_0 = 65~{\rm km}~{\rm s}^{-1}~{\rm Mpc}^{-1}$ gives a distance to the SN of $\sim21.6~{\rm Mpc}$. We note, however, that NGC~5033 is the host of several historical supernovae, including the radio emitting SN 1985L \citep{VanDyk98}. If we follow \citet{VanDyk98} and estimate the \citet{Tully88} distance with Hubble constant $H_0 = 70~{\rm km}~{\rm s}^{-1}~{\rm Mpc}^{-1}$, we obtain  a distance to NGC~5033 of $\sim20~{\rm Mpc}$. Since a change in the distance estimate of $<10\%$ will not significantly affect any of our estimates of the SN physical properties, we will, for consistency with the previous radio work, use in this paper the Van Dyk distance of 20 Mpc. However, it should be noted that \citet{PerezTorres05} assumed a much shorter distance of 13.1 Mpc.

SN 2001gd was first detected at radio wavelengths on 2002 February 8.54 by \citet{Stockdale02}, using the Very Large Array (VLA)\footnote{The VLA telescope of the National Radio Astronomy Observatory is operated by Associated Universities, Inc. under a cooperative agreement with the National Science Foundation} at 1.3 cm (22.4 GHz), 3.6 cm (8.4 GHz), 6 cm (4.9 GHz), and 20 cm (1.5 GHz).  They report a radio position of $\alpha = 13^h 13^m 23\fs889$, $\delta = +36\arcdeg 38\arcmin 18\farcs14$ (J2000) with uncertainties of $\pm 0\farcs2$ in each coordinate, which agrees with the reported optical position of \citet{Nakano01} to within the errors.  

\citet{Stockdale03} presented the results of initial radio monitoring of SN 2001gd between 2002 February 8 and 2002 October 28. Their results showed that the radio emission evolves very smoothly, consistent with the nonthermal-emitting, thermal-absorbing model described by \citet{Weiler02}.  Their results also indicate a CSM with significant structure embedded in a uniform medium, similar to that proposed for Type IIb SN 1993J \citep{VanDyk94,Weiler07}.  The similar optical spectra and early radio evolution of both SN 1993J and SN 2001gd indicate that the two Type IIb SNe may have a number of similarities.

\citet{PerezTorres05} conducted VLBI observations of SN 2001gd at 3.6 cm (8.4 GHz) on 2002 June 26 and 2003 April 8.  Although they were unable to resolve any structure in the SN, they were able to estimate the SN angular size through interferometer visibility fitting and found that their data suggested a possible deceleration in the expansion (although their data were not conclusive).  Using our assumed distance of 20 Mpc to the host galaxy, and the angular size limits estimated by \citet{PerezTorres05}, the upper limit to the expansion velocity of the blastwave is $\lesssim22,000$ \kms\ for the 2002 June 26 VLBI observation (day 296).  Depending on the exact nature of the emitting region, the upper limit may be as low as $\lesssim19,000$ \kms.  Thus, for the purposes of this paper, and for comparison with SN~1993J, we assume an expansion velocity of 15,000 \kms.  

\citet{PerezTorres05} conducted multifrequency radio observations with the VLA on 2003 April 8 to complement their VLBI data, obtaining a spectrum consistent with synchrotron emission partially suppressed by free-free absorption.  They also obtained X-ray data on SN 2001gd from XMM-Newton observations of NGC5033 on 2001 July 2 and 2002 December 18 and report the detection of X-ray emission on the latter date consistent with the expected reverse shock emission from typical Type II SNe.  

Recent VLA observations indicate a dramatic change in the evolution of the radio emission from SN~2001gd.  These new observations and their implications are presented in this paper.

\section{Radio Observations}

\begin{figure}
\epsscale{0.75}
\includegraphics[angle=0,scale=0.65]{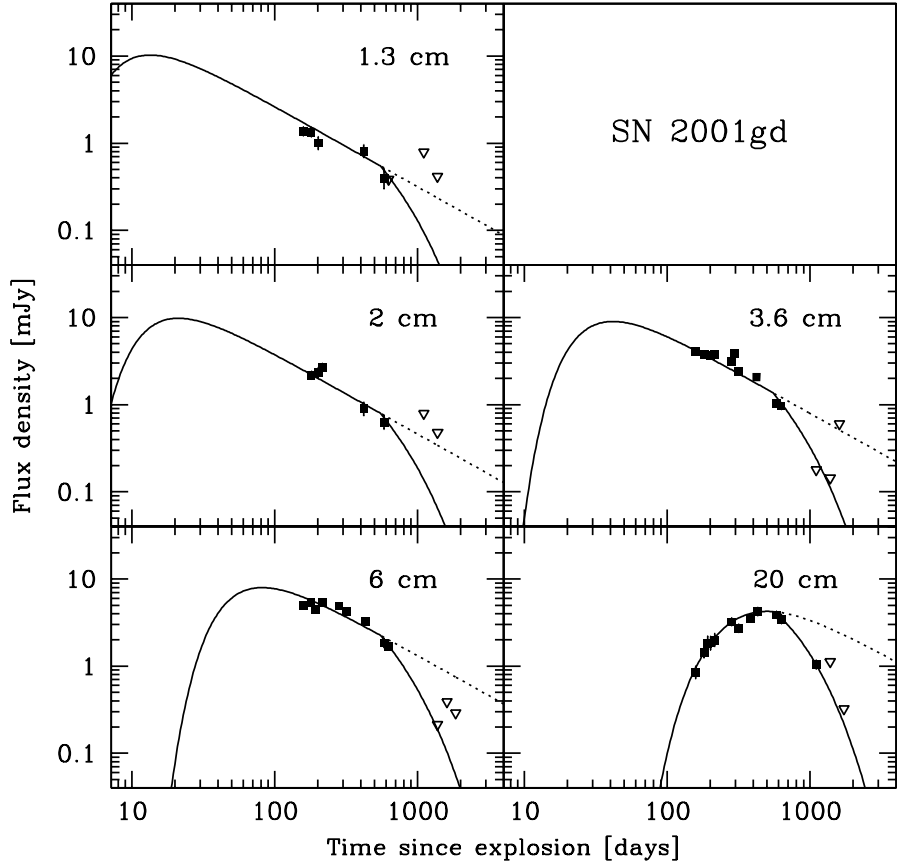}
\caption{Radio ``light curves'' for SN 2001gd in NGC 5033.  The five wavelengths, 1.3 cm (22.4 GHz; {\it top left}), 2 cm (14.9 GHz; {\it middle left}), 3.6 cm (8.4 GHz; {\it middle right}), 6 cm (4.9 GHz; {\it bottom left}), and 20 cm (1.5 GHz; {\it bottom right}) are shown together with their best-fit model light curves.  The SN age is in days since the adopted explosion date of 2001 September 03.  Because the decline index $\beta$ of the radio emission steepened after day $\sim550$, following the example of SN~1993J \citep{Weiler07}, we have fitted the ``late'' data after day 550 with an exponential decline with an e-folding time of 500 days. The inverted open triangles represent $3\sigma$ upper limits and the dotted lines show the extrapolation of the fit to the ``early" data before day $\sim550$ given in Table \ref{tab3}}
\label{fig1}
\end{figure}

New radio observations of SN 2001gd have been made with the VLA at 1.3 cm (22.4 GHz), 2 cm (14.9 GHz), 3.6 cm (8.4 GHz), 6 cm (4.9 GHz), and 20 cm (1.5 GHz), with one upper limit at 90 cm (0.3 GHz) from 2002 November 08 through 2006 September 25 and are presented in Table~\ref{tab1} and Figure~\ref{fig1}.  
For completeness and ease of reference, we include the previously published results from \citet{Stockdale03}, and \citet{PerezTorres05}, as well as one measurement from the GMRT \citep{Chandra02}, in both Table~\ref{tab1} and Figure~\ref{fig1}.

\begin{figure}
\epsscale{0.60}
\includegraphics[angle=0,scale=0.65]{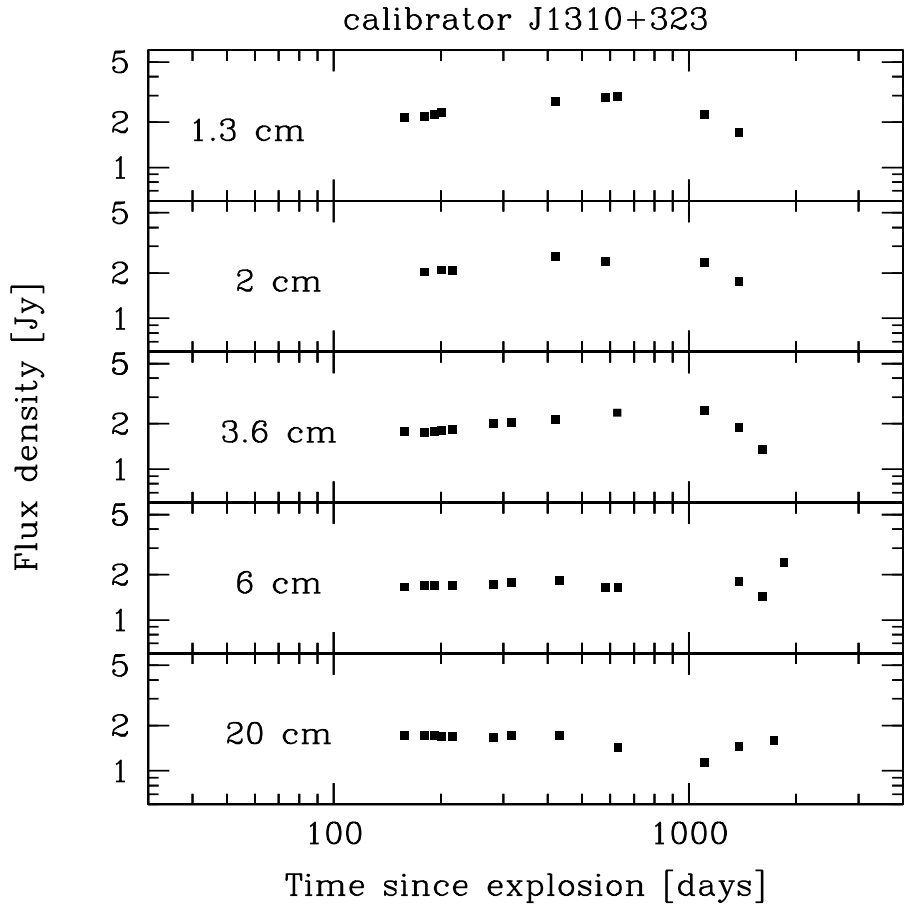}
\caption{ Flux density measurements of the secondary calibrator J1310+323 at the five wavelengths of 1.3 cm (22.4 GHz; {\it top}), 2 cm (14.9 GHz; {\it one below top}), 3.6 cm (8.4 GHz; {\it middle}), 6 cm (4.9 GHz; {\it one above bottom}), and 20 cm (1.5 GHz; {\it bottom}).}
\label{fig2}
\end{figure}

The techniques of observation, editing, calibration, and error estimation are described in previous publications on the radio emission from SNe \citep[see, \eg][]{Weiler86,Weiler02,Sramek03}.  We used 3C286 as the primary calibrator for flux density bootstrapping, while J1310+323, with a defined position of  $\alpha = 13^{\rm h}10^{\rm m}28{\fs}66$, $\delta=+32^{\circ}20\arcmin43{\farcs}8$ served  as the phase calibrator and secondary flux density calibrator.  The flux densities of J1310+323 are shown in Table~\ref{tab2} and plotted in Figure~\ref{fig2}.  Like most phase calibrators, J1310+323 shows some variation in its flux density because phase calibrators are selected for their proximity to the SN and unresolvable angular size rather than their flux stability. The primary calibrator is therefore used to determine the absolute flux density of the secondary calibrator at each epoch of the observations.

The listed flux density measurement errors are determined based on both the intrinsic RMS background noise level of the particular radio map, and a fractional error, $\epsilon$, proportional to the SN flux.  The RMS map noise results from both small unresolved fluctuations in the background emission and random map fluctuations due to receiver noise, while the fractional error, $\epsilon$, accounts for the inaccuracy of VLA flux density calibration \citep[see, \eg][]{Weiler86} and possible deviations of the primary calibrator from an absolute flux density scale.  The final errors ($\sigma_f$) given for the measurements of SN 2001gd are taken as the addition in quadrature of these components:

\begin{equation}
\label{eq1}
\sigma_{f}^{2} = (\epsilon S_0)^2+\sigma_{0}^2 
\end{equation} 

\noindent where $S_0$ is the measured flux density, $\sigma_0$ is the map RMS
background noise for each
observation, and $\epsilon = 0.15$ for 90 cm, $\epsilon =0.1$ for 20 cm, 
$\epsilon =0.05$ for 6 cm and 3.6 cm, $\epsilon =0.075$ for 2 cm, and 
$\epsilon =0.1$ for 1.3 cm. 

\section{Parameterized Radio Light Curves\label{lightcurve}}

From the most recent RSN modeling discussion of \citet{Weiler02}
and \citet{Sramek03}, we use a parameterized model :

\begin{equation}
\label{eq2}
S(\mbox{mJy}) = K_1 \left(\frac{\nu}{\mbox{5\
GHz}}\right)^{\alpha} \left(\frac{t-t_0}{\mbox{1\ day}}\right)^{\beta}
e^{-\tau_{\rm external}} \left(\frac{1-e^{-\tau_{{\rm CSM}_{\rm clumps}}}}{\tau_{{\rm CSM}_{\rm
clumps}}}\right) \left(\frac{1-e^{-\tau_{\rm internal}}}{\tau_{\rm internal}}\right) 
\end{equation} 

\noindent with  

\begin{equation}
\label{eq3}
\tau_{\rm external}  =  \tau_{{\rm CSM}_{\rm homogeneous}}+\tau_{\rm distant},
\end{equation}

\noindent where

\begin{equation}
\label{eq4}
\tau_{{\rm CSM}_{\rm homogeneous}}  =  K_2
\left(\frac{\nu}{\mbox{5 GHz}}\right)^{-2.1}
\left(\frac{t-t_0}{\mbox{1\ day}}\right)^{\delta}
\end{equation}

\begin{equation}
\label{eq5}
\tau_{\rm distant}  =  K_4  \left(\frac{\nu}{\mbox{5\
GHz}}\right)^{-2.1}
\end{equation} 

\noindent and

\begin{equation}
\label{eq6}
\tau_{{\rm CSM}_{\rm clumps}}  =  K_3 \left(\frac{\nu}{\mbox{5\
GHz}}\right)^{-2.1} \left(\frac{t-t_0}{\mbox{1\
day}}\right)^{\delta^{\prime}}
\end{equation} 

\noindent with $K_1$, $K_2$, $K_3$, and $K_4$ determined from fits to
the data and corresponding to the flux density ($K_1$), homogeneous ($K_2$, $K_4$), and clumpy or filamentary ($K_3$) free-free absorption (FFA) at 5~GHz, formally but not necessarily physically, one day after the explosion date $t_0$. For a more detailed explanation of the physical meaning of the parameters, the reader should consult the references cited above.

Since it is physically realistic and may be needed in some RSNe where
radio observations have been obtained at early times and high
frequencies, Equation (\ref{eq2}) also includes the possibility for an
internal absorption term.  This internal absorption 
($\tau_{\rm internal}$) term may consist of two parts -- synchrotron
self-absorption (SSA; $\tau_{{\rm internal}_{\rm SSA}}$), and mixed,
thermal FFA/non-thermal emission ($\tau_{{\rm internal}_{\rm FFA}}$). 

\begin{equation}
\label{eq8}
\tau_{\rm internal}  = \tau_{\rm internal_{\rm SSA}} + \tau_{\rm internal_{\rm FFA}}
\end{equation}

\begin{equation}
\label{eq9}
\tau_{\rm internal_{\rm SSA}} = K_5\left(\frac{\nu}{\mbox{5\
GHz}}\right)^{\alpha-2.5}  \left(\frac{t-t_0}{\mbox{1\
day}}\right)^{\delta^{\prime\prime}}
\end{equation}

\begin{equation}
\label{eq10}
\tau_{\rm internal_{\rm FFA}}  =   K_6  \left(\frac{\nu}{\mbox{5\
GHz}}\right)^{-2.1} \left(\frac{t-t_0}{\mbox{1\
day}}\right)^{\delta^{\prime\prime\prime}}
\end{equation}

\noindent with $K_5$ corresponding to the internal, non-thermal ($\nu^{\alpha - 2.5}$) SSA and $K_6$ corresponding to the internal thermal ($\nu^{-2.1}$) free-free absorption mixed with non-thermal emission, at 5~GHz, formally but not necessarily physically, one day after the explosion date $t_0$. 
The parameters $\delta^{\prime \prime}$ and $\delta^{\prime \prime
\prime}$ describe the time dependence of the optical depths for the SSA
and FFA internal absorption components, respectively. 

Application of this basic parameterization has been shown to be effective
in describing the physical characteristics of the presupernova system,
its CSM, and its final stages of evolution before explosion for objects
ranging from the two decades of monitoring the complex radio emission
from SN 1979C \citep{Montes00} through the unusual SN 1998bw (GRB980425)
\citep{Weiler01} to the very well studied SN~1993J \citep{Weiler07}.

\section{Radio Light Curve Description\label{RLCD}}

The multi-frequency radio data for SN 2001gd are listed in Table~\ref{tab1} and displayed in Figure~\ref{fig1}.  Examination of the data indicates that radio evolution of the SN is best described by two distinct time intervals: an ``early" interval before day $\sim550$ and a ``late" interval thereafter.  Although the precise date of the observed ``break" is not well defined, it appears that around the time of the observations of \citet{PerezTorres05} on 2003 April 8 (day 582) the rate of flux decline markedly steepened such that the data can no longer be adequately described by the model appropriate for earlier times.  Similar ``breaks" have been observed in the radio light curves of other RSNe, including SN~1988Z \citep{Williams02}, SN 1980K \citep{Montes98}, and SN~1993J \citep{Weiler07}.

An extensive search for the best parameter fits to equations (\ref{eq2})-(\ref{eq10}) was carried out on the ``early'' interval (day $<550$) by minimizing the reduced $\chi^2$ ($\chi^2_{\rm red}$).  Since there are only four radio detections after the measurements on day 582, the ``late'' radio light curve behavior is very poorly defined except to note that the few detections and the several upper limits indicate that the decline rate has significantly steepened and, following the example of SN~1993J \citep{Weiler07}, are consistent with an exponential decline with an e-folding time of 500 days.

The early epoch fitting parameters are consistent with a homogeneous ($K_2$) CSM, although more complex structures such a CSM containing filamentary ($K_3$) components cannot be ruled out. No evidence is seen for any distant, homogeneous thermal absorption ($K_4=0$), and the very early data ($<100$ days) are too sparse to determine from the fitting if SSA ($K_5$), or mixed, internal, free-free absorption/nonthermal emission ($K_6$) may have been present.  

However, we have estimated the brightness temperatures for the ``early'' (before day $\sim550$) 20~cm emission based on a pure FFA model and found that they exceed the limiting temperature for synchrotron emission of $\sim10^{11.5}$ K calculated by \citet{Readhead94}, which was also the case for SN~1993J \citep{Weiler07}.  Given our inability to model this early absorption based purely on our sparse ``early'' data, we have chosen to assume that SN~2001gd has a similar brightness temperature evolution ($\sim10^{11}$ K at peak) for the 20~cm emission as it becomes optically thin. Given that SN~2001gd and SN~1993J have similar characteristics at ``early'' times, we estimate, as was the case for SN~1993J, that a significant SSA component is required for SN~2001gd along with the FFA component determined from fitting which describes the rapid rise of the 20~cm data well. The best parameters which we can estimate for the ``early'' epoch of SN~2001gd with these assumptions are given in Table~\ref{tab3}. We note an apparent similarity between the $\delta$ and $\delta^{\prime\prime}$ terms from our modeling of the SN~2001gd radio emission. This is similar to the case of SN~1993J where the two values were $\delta=-1.88$ and $\delta^{\prime\prime} = -2.05$.  However, it must be stated that, from pure modeling, our sparse ``early'' data only constrain the values of  $\delta$ and $\delta^{\prime\prime}$ for SN~2001gd to lie between $-1$ and $-2$. To narrow the range of these values, we have therefore assumed a value of $\delta = -1.88$ to match the value for $\delta$ determined for SN~1993J \citep{Weiler07}. 

The ``late'' period appears to be optically thin at all observed frequencies with a much steeper decline rate than is possible with the ``early'' data determined $\beta = -0.92$ (shown as the dashed line in Figure~\ref{fig1}). Again, if we assume a similarity to the exponential decline seen for SN~1993J \citep{Weiler07}, the late epoch of SN~2001gd is consistent with an exponential decline after day 550, with an e-folding time of 500 days.  This observed ``break" in the radio light curves of SN~2001gd is interpreted as indicative of a rapid transition of the blastwave into a much more tenuous medium.  This occurs much earlier for SN~2001gd (day $\sim550$) than that seen for SN~1993J (day $\sim3,100$) by \cite{Weiler07}.

Before the break, SN~2001gd is still somewhat optically thick at 20~cm, but after the break it appears to be optically thin.  This discounts the possibility of the break being the result of a sudden change in the shock velocity or magnetic field properties at the shock front since such changes would not yield an optical depth reduction. Given the similar expansion velocities of $\lesssim19,000$ \kms\ for SN~2001gd (derived from \citep{PerezTorres05}; see the Introduction) and $11,000 - 15,000$ \kms\ for SN~1993J \citep{Marcaide97,Marcaide07}, it is clear that both supernovae had a period of enhanced mass-loss rate preceding their explosions.  Thus, assuming a blastwave velocity for SN~2001gd of $15,000$ \kms\ and a ``standard'' pre-explosion wind velocity of 10 \kms, this enhanced mass-loss persisted for the $\sim1,800$ years prior to the explosion.  However in the case of SN~1993J, this phase began $\sim$ 8,800 years earlier than the explosion of its progenitor star. The shorter duration of the high presupernova mass-loss rate for SN~2001gd is also borne out in its faster rise and higher luminosity, indicating there was less intervening, presupernova wind established, CSM to contribute to the FFA at early epochs. 

\subsection{The Mass-Loss Rate for the SN Progenitor\label{MLprogen}}

Through use of this modeling, a number of physical properties of SNe can be determined from the radio observations.  One of these is the mass-loss rate from the SN progenitor prior to explosion. From the Chevalier (1982a,b) model, the turn-on of the radio emission for RSNe provides a measure of the presupernova mass-loss rate to wind velocity ratio ($\dot M/w_{\rm wind}$) and 
\citet{Weiler86} derived this ratio for the case of pure, thermal, external absorption by a homogeneous medium.  

Given the limited amount of early radio
data available for SN 2001gd, there is no need to invoke more complex
models than the simple, uniform, external thermal absorption 
model, so that $\tau(t=1~day)=K_2$ and  equation (16) of
\citet{Weiler86} can be written:

\begin{eqnarray}
\label{eq11}
\frac{\dot M (M_\odot\ {\rm yr}^{-1})}{( w_{\rm wind} / 10\ {\rm km\
s}^{-1} )} & = & 3.0 \times 10^{-6}\ \phi \ \tau_{{\rm CSM}_{\rm homogeneous}}^{0.5} \ m^{-1.5}
{\left(\frac{v_{\rm i}}{10^{4}\ {\rm km\ s}^{-1}}\right)}^{1.5} \times
\nonumber \\ & & {\left(\frac{t_{\rm i}}{45\ {\rm days}}\right) }^{1.5}
{\left(\frac{t}{t_{\rm i}}\right) }^{1.5 m}{\left(\frac{T}{10^{4}\ {\rm
K}} \right)}^{0.68} .
\end{eqnarray}

\noindent Here, the extra factor $\phi$ is a small correction that takes
into account the fact that, for SN~2001gd, the CSM density behaves like
$\rho \propto r^{-1.61}$ (see below) instead of  $r^{-2}$ as it would under
the usual constant mass-loss rate assumption. The factor  $\phi$ is given by the square root of the ratio of the integration constant for $\tau$  in the case of $\rho \propto r^{-1.61}$ to the one appropriate for $\rho \propto  r^{-2}$ , i.e.

\begin{equation}
\label{eq12}
\phi \ = \ \left(\frac{2\times 1.61-1}{2\times 2 - 1}\right)^{0.5} \ = \ 0.86
\end{equation}


Since SN 2001gd is a Type IIb supernova as SN 1993J was,  and
appears to be quite similar to it, we will adopt for SN 2001gd some of the
same parameters as measured for SN 1993J. Thus, for SN 2001gd we assume
$v_{\rm i} = 15,000$ \kms\ at $t= 45$ days, which is a value consistent
with the results of \citet{PerezTorres05} as discussed earlier.

We also adopt values of $T = 20,000$ K, $w_{\rm wind} = 10$ \kms\ (which
is appropriate for a RSG wind), $t = (t_{\rm 6 cm\ peak} - t_0)$ days
from our best fits to the radio data, and $m=0.845$, as measured for SN
1993J by \cite{Marcaide07}.   With the assumptions for the blastwave and
CSM properties discussed above, and the results for the best-fit
parameters listed in Table \ref{tab3}, our estimated presupernova
mass-loss rate is $\dot M = 5.0\times 10^{-6}\ M_\odot$ yr$^{-1}$ at 8.6 years before explosion.  

The fitting parameter value of $\delta = -1.88$ indicates that  the CSM density behaves like $\rho \propto  r^{-1.61}$ so that the mass-loss rate for SN 2001gd
was not constant but was decreasing in the years leading up to the
explosion, i.e. $\dot M \propto r^2 \rho w_{\rm wind} \propto r^{0.39}$ for
a constant  $w_{\rm wind}$.  

Since the shock speed for SN~2001gd is assumed to be $\sim15,000$ \kms\
at $t=45$ days, but variable with time ($v\propto t^{-0.155}$ for $m=0.845$ from \citealt{Marcaide07}), and the pre-supernova  wind velocity for an RSG is typically $\sim 10$ \kms, we calculate that the abrupt decline in the radio light curves starting around day $\sim550$ implies this change took place in the presupernova stellar wind $\sim1,800$ years
before explosion when the mass-loss rate was as high as $3.9\times 10^{-5}\ M_\odot$ yr$^{-1}$. Integrating the mass-loss rate over the last $\sim1,800$ years before explosion, we find that, during that time, the progenitor star shed $\sim 0.05~M_\odot$ in a massive stellar wind. At earlier epochs of the progenitor's evolution the mass-loss rate was considerably lower, as indicated by the radio light curve ``break'' discussed above and the transition of the shock wave to a lower density CSM at that time. 

\section{Conclusions}

Figure 1 and Table 3 show that the radio light curves
for SN 2001gd can be described by standard RSN models
\citep{Weiler86,Weiler90,Weiler02,Montes97} during its ``early''
evolution up to day $\sim550$. After day $\sim550$ the flux density
decline rate clearly steepened  to, in comparison with SN~1993J
\citep{Weiler07}, a form consistent with an exponential decay with
e-folding time of 500 days. The ``break'' in the radio light curves for
SN 2001gd is interpreted as due to a change in the average CSM density.
Similar ``breaks'' have been identified earlier in SN~1980K
\citep{Montes98}, SN~1988Z \citep{Williams02}, and SN~1993J
\citep{Weiler07}.

Normal assumptions for presupernova stellar wind properties imply a
dense wind with mass-loss rate decreasing for the last $\sim1,800$ years
before explosion from $\dot M \sim3.9 \times 10^{-5}\ M_\odot$ yr$^{-1}$ to $\dot M \sim 5.0\times 10^{-6}\ M_\odot$ yr$^{-1}$ around the time of explosion. The mass-loss rate was sharply lower for times earlier than $\sim1,800$ years
before explosion.

\acknowledgments

KWW wishes to thank the Office of Naval Research (ONR) for the 6.1 funding supporting his research.  CJS is a Cottrell Scholar of Research Corporation and work on this project has been supported by the NASA Wisconsin Space Grant Consortium. NP is Astronomer Emeritus at the Space Telescope Science Institute (STScI) that kindly provided research facilities and partial support for this work. Additional information and data on radio supernovae can be found on {\it http://rsd-www.nrl.navy.mil/7213/weiler/sne-home.html}
and linked pages.

\clearpage

\begin{deluxetable}{ccccccccc}
\tablecolumns{9}
\tablewidth{6.35in}
\tabletypesize{\scriptsize}
\tablecaption{Flux Density Measurements for SN 2001gd\tablenotemark{a}}
\tablehead{
\colhead{Obs.~Date} & \colhead{Age} & \colhead{Tel.} & \colhead{$S$ (90
cm)} & \colhead{$S$ (20 cm)} & \colhead{$S$ (6 cm)} 
& \colhead{$S$ (3.6 cm)} & \colhead{$S$ (2 cm)} & \colhead{$S$ (1.3 cm)}\\
\colhead{(UT)} & \colhead{(days)} &  &
\colhead{(mJy)} & \colhead {(mJy)} & \colhead {(mJy)} & \colhead {(mJy)}
& \colhead{(mJy)} & \colhead{(mJy)}} 
\startdata
2001 Sep 03  & $\equiv$ 0 & & & & \\
2002 Feb 08 & 159 & VLA-A &\nodata & $0.85 \pm 0.15$ & $5.05 \pm 0.26$ & $4.09 \pm 0.21$ & \nodata & $1.37 \pm 0.18$ \\
2002 Mar 02 & 180 & VLA-A &\nodata & $1.41 \pm 0.20$ & $5.41 \pm 0.28$ & $3.73 \pm 0.20$ & $2.18 \pm 0.24$ & $1.34 \pm 0.18$ \\
2002 Mar 14 & 192 & VLA-A &\nodata & $1.84 \pm 0.37$ & $ 4.46 \pm 0.24 $ & \nodata & \nodata & \nodata\\
2002 Mar 23 & 201 & VLA-A &\nodata & $ 1.83 \pm 0.36 $ & \nodata & $ 3.66 \pm 0.21 $ & $ 2.33 \pm 0.32 $ & $ 1.00 \pm 0.19 $\\
2002 Apr 06 & 215 & VLA-A &\nodata & $ 1.99 \pm 0.37 $ & $ 5.33 \pm 0.28 $ & $ 3.78 \pm 0.21 $ & $ 2.66 \pm 0.30 $ & \nodata\\
2002 Jun 12 & 282 & VLA-B &\nodata & $ 3.23 \pm 0.37 $ & $ 4.84 \pm 0.25 $ & $ 3.10 \pm 0.17 $ & \nodata & \nodata\\
2002 Jun 26\tablenotemark{b} & 296 & VLBA  &\nodata & \nodata & \nodata & $ 3.83 \pm .19 $&\nodata & \nodata \\
2002 Jul 16 & 317 & VLA-B &\nodata & $ 2.73 \pm 0.31 $ & $ 4.26 \pm 0.23 $ & $ 2.39 \pm 0.14 $ & \nodata & \nodata\\
2002 Sep 22\tablenotemark{c} & 384 & GMRT & \nodata & $3.55 \pm 0.22$& \nodata & \nodata & \nodata & \nodata \\
2002 Oct 28 & 421 & VLA-C &\nodata & \nodata & \nodata & $ 2.07 \pm 0.14 $ & $ 0.89 \pm 0.18 $ & $ 0.81 \pm 0.16 $\\
2002 Nov 08 & 432 & VLA-C & \nodata & $4.22 \pm 0.48$ & $3.26 \pm 0.20$ & \nodata & \nodata & \nodata \\
2003 Apr 08\tablenotemark{b} & 582 & VLA-D &\nodata & $ 3.89 \pm 0.40 $ & $ 1.85 \pm 0.24 $ & $ 1.02 \pm 0.05 $ & $ 0.63 \pm 0.13 $ & $ 0.39 \pm 0.12$ \\ 
2003 May 23 & 627 & VLA-A & $<3.09$  & \nodata & \nodata & $ 0.96 \pm 0.10 $ & \nodata & $ <0.39 $\\
2003 May 26 & 630 & VLA-A & \nodata & $3.47 \pm 0.45$ & $1.67 \pm 0.10$ & \nodata & \nodata & \nodata \\
2004 Sep 10 & 1104 & VLA-A &\nodata & $ 1.04 \pm 0.14 $ & \nodata & $ <0.18 $ & $ <0.79 $ & $ <0.79 $\\
2005 Jun 14 & 1380 & VLA-C &\nodata & $ <1.12 $ & $ <0.21 $ & $ <0.14 $ & $ <0.48 $ & $ <0.42 $\\
2006 Jan 24 & 1605 & VLA-D & \nodata & \nodata & $<0.39$ & $<0.6$ & \nodata & \nodata \\
2006 May 29 & 1729 & VLA-BnA & \nodata & $<0.32$ & \nodata & \nodata & \nodata & \nodata \\
2006 Sep 25 & 1849 & VLA-B & \nodata & \nodata & $<0.29$ &  \nodata & \nodata & \nodata \\\enddata
\tablenotetext{a}{All upper limits are 3$\sigma$.}
\tablenotetext{b}{From \citet{PerezTorres05}; \citet{PerezTorres05} also measured a 3$\sigma$ upper limit of $<0.3$ mJy for SN 2001gd on 2003 Apr 08 at 0.7 cm (43.3 GHz).}
\tablenotetext{c}{\citet{Chandra02}}  
\label{tab1}
\end{deluxetable}

\clearpage

\begin{deluxetable}{ccccccccc}
\tablecolumns{9}
\tablewidth{6.0in}
\tabletypesize{\scriptsize}
\tablecaption{Flux Density Measurements for Secondary Calibrator J1310+323}
\tablehead{
\colhead{Obs.~Date} & \colhead{Age} & \colhead{Tel.} & \colhead{$S$ (90
cm)} & \colhead{$S$ (20 cm)} & \colhead{$S$ (6 cm)} 
& \colhead{$S$ (3.6 cm)} & \colhead{$S$ (2 cm)} & \colhead{$S$ (1.3 cm)}\\
\colhead{(UT)} & \colhead{(days)} &
\colhead{(Jy)} & \colhead {(Jy)} & \colhead {(Jy)} & \colhead {(Jy)}
& \colhead{(Jy)} & \colhead{(Jy)}} 
\startdata
2002 Feb 08 & 159 & VLA-A & \nodata & 1.70 & 1.66 & 1.77 & \nodata & 2.15 \\
2002 Mar 02 & 180 & VLA-A & \nodata & 1.70\tablenotemark{a} & 1.69 & 1.75 & 2.03 & 2.16 \\
2002 Mar 14 & 192 & VLA-A & \nodata & 1.70\tablenotemark{a} & 1.70\tablenotemark{a} & \nodata & \nodata & \nodata \\
2002 Mar 23 & 201 & VLA-A & \nodata & 1.68 & \nodata & 1.80\tablenotemark{a} & 2.08 & 2.32 \\
2002 Apr 06 & 215 & VLA-A & \nodata & 1.69 & 1.70\tablenotemark{a} & 1.83\tablenotemark{a} & 2.07 & \nodata \\
2002 Jun 12 & 282 & VLA-B & \nodata & 1.67 & 1.73 & 2.01 & \nodata & \nodata \\
2002 Jul 16 & 296 & VLA-B & \nodata & 1.72 & 1.77 & 2.05 & \nodata & \nodata \\
2002 Oct 28 & 421 & VLA-C & \nodata & \nodata & \nodata & 2.13 & 2.55 & 2.73 \\
2002 Nov 08 & 432 & VLA-C & \nodata & 1.70 & 1.83 & \nodata & \nodata & \nodata \\ 
2003 Apr 08\tablenotemark{b} & 582 & VLA-D & \nodata & 1.54 & 1.66 & \nodata & 2.40 & 2.91 \\
2003 May 23 & 627 & VLA-A & 2.74 & \nodata & \nodata & 2.37 & \nodata & 2.93 \\
2003 May 26 & 630 & VLA-A & \nodata & 1.42 & 1.65 & \nodata & \nodata & \nodata \\ 
2004 Sep 10 & 1104 & VLA-A & \nodata & 1.14 & \nodata & 2.47 & 2.35 & 2.24 \\
2005 Jun 14 & 1380 & VLA-C & \nodata & 1.45 & 1.81 & 1.89 & 1.76 & 1.71 \\
2006 Jan 24 & 1605 & VLA-D &\nodata & \nodata  & 1.44 & 1.35 & \nodata & \nodata \\
2006 May 29 & 1729 & VLA-BnA &\nodata & 1.58  & \nodata & \nodata & \nodata & \nodata \\
2006 Sep 25 & 1849 & VLA-B &\nodata & \nodata  & 2.42 & \nodata & \nodata & \nodata \\
\enddata
\tablenotetext{a}{Value was not measured but was interpolated from the closest calibration at that frequency.}
\tablenotetext{b}{From \citet{PerezTorres05}; \citet{PerezTorres05} also measured a value of 2.58 Jy for J1310+323 on 2003 Apr 08 at 0.7 cm (43.3 GHz).}
\label{tab2}
\end{deluxetable}

\clearpage

\begin{deluxetable}{lc}
\tablecolumns{2}
\tablewidth{3.5in}
\tablecaption{Fitting Parameters for SN 2001gd} 
\tablehead{
\colhead{Parameter} & \colhead{Early ({\rm day} $<550$)} 
}
\startdata
$K_1$ & $7.5 \times 10^2$ \\
$\alpha$ & $-0.94$  \\
$\beta$  & $-0.92$ \\
$K_2$ & $1.5 \times 10^3$ \\
$\delta$ & $-1.88$\\
$K_5$ & $1.7 \times 10^2$ \\
$\delta^{\prime}$ & $-1.5$ \\
$t_0$\tablenotemark{a} & 2001 Sep 03 \\
${\rm S_{6 cm\ peak}}$ (mJy) & 7.96 \\
${\rm L_{6 cm\ peak}}$ (erg s$^{-1}$ Hz$^{-1}$) & $3.8 \times 10^{27}$ \\
$t = (t_{\rm 6\ cm\ peak} - t_0)$ (days) & 80 \\ 
$\dot M$ ($M_\odot$ yr$^{-1}$)\tablenotemark{b} & $(0.5-3.9) \times 10^{-5}$\\
${\rm Assumed \ distance}$ (Mpc) & 20 \\
\enddata
\tablenotetext{a}{Defined in the Introduction.}
\tablenotetext{b}{Assuming $w_{\rm wind}= 10~{\rm km}~{\rm s}^{-1}$, 
$v_{\rm i}=v_{\rm blastwave} = 15,000~\rm{km}~\rm{s}^{-1}$, 
$\rm{T}=20,000~\rm{K}$, and ${\rm t_i} = 45~{\rm days}$.}
\label{tab3}
\end{deluxetable}

\clearpage

\clearpage

\begin{thebibliography}{}
\bibitem[Chandra et al. (2002)]{Chandra02} Chandra, P., Ray, A., \& Bhatnagar, S.\ 2002, \iaucirc~7982, 2 
\bibitem[Chevalier (1982a)]{Chevalier82a} Chevalier, R.\ A. 1982a, \apj, 259, 302
\bibitem[Chevalier (1982b)]{Chevalier82b} Chevalier, R.\ A. 1982b, \apjl, 259, L85
\bibitem[Marcaide et al.(1997)]{Marcaide97} Marcaide, J.~M. \etal\ 1997, \apjl, 486, L31
\bibitem[Marcaide \etal (2007)]{Marcaide07} Marcaide, J.~M. \etal\ 2007, \mnras, submitted
\bibitem[Matheson \etal (2001)]{Matheson01} Matheson, T., Jha, S., Challis, P., Kirshner, R., \& Berlind, P. 2001, \iaucirc~7765
\bibitem[Montes, Weiler, \& Panagia(1997)]{Montes97} Montes, M.\ J., Weiler, K.\ W., \& Panagia, N.\ 1997, \apj, 488, 792 
\bibitem[Montes \etal (1998)]{Montes98} Montes, M.~J., van Dyk, S.~D., Weiler, K.~W., Sramek, R.~A., \& Panagia, N.\ 1998, \apj, 506, 874 
\bibitem[Montes \etal (2000)]{Montes00} Montes, M.~J., Weiler, K.~W., Van Dyk, S.~D., Panagia, N., Lacey, C.~K., Sramek, R.~A., \& Park, R.\ 2000, \apj, 532, 1124 
\bibitem[Nakano \etal (2001)]{Nakano01} Nakano, S., Itagaki, K., Kushida, Y., Kushida, R., \& Dimai, A. 2001, \iaucirc\ 7761
\bibitem[Natta \& Panagia (1984)]{Natta84} Natta, A., \& Panagia, N. 1984, \apj, 287, 228
\bibitem[Osterbrock (1974)]{Osterbrock74}  Osterbrock, D.~E.~1974, Astrophysics of Gaseous Nebulae (Freeman, San Francisco), p. 82
\bibitem[P{\'e}rez-Torres \etal (2005)]{PerezTorres05} P{\'e}rez-Torres, M.~A., et al.\ 2005, \mnras, 360, 1055 
\bibitem[Readhead (1994)]{Readhead94} Readhead, A.~C.~S. 1994, \apj, 426, 51
\bibitem[Sramek \& Weiler (2003)]{Sramek03} Sramek, R.~A. \& Weiler, K.~W.\ 2003, Springer Lecture Notes in Physics Vol.~598: Supernovae and Gamma-Ray Bursters, 598,  137
\bibitem[Stockdale \etal (2002)]{Stockdale02} Stockdale, C. J., Perez-Torres, M. A., Marcaide, J. M., Sramek, R. A., Weiler, K. W., Van Dyk, S. D., Panagia, N., Lundqvist, P., Pooley, D., Immler, S., \& Lewin, W. 2002, \iaucirc\ 7830
\bibitem[Stockdale \etal (2003)]{Stockdale03} Stockdale, C.~J., Weiler, K.~W., Van Dyk, S.~D., Montes, M.~J., Panagia, N., Sramek, R.~A., Perez-Torres, M.~A., \& Marcaide, J.~M.\ 2003, \apj, 592, 900 
\bibitem[Tully(1988)]{Tully88} Tully, R.~B.\ 1988, Nearby Galaxies Catalog (Cambridge and New York: Cambridge University Press)
\bibitem[Van Dyk \etal (1994)]{VanDyk94} Van Dyk, S., Weiler, K., Sramek, R., Rupen, M., \& Panagia, N.\ 1994, \apjl, 432, 115
\bibitem[Van Dyk \etal (1998)]{VanDyk98} Van Dyk, S.~D., Montes, M.~J., Weiler, K.~W., Sramek, R.~A., \& Panagia, N.\ 1998, \aj, 115, 1103 
\bibitem[Weiler \etal (1986)]{Weiler86}  Weiler, K.~W., Sramek, R.~A., Panagia, N., van der Hulst, J.~M., \& Salvati, M. 1986, \apj, 301, 790
\bibitem[Weiler \etal (1990)]{Weiler90}  Weiler, K.~W., Panagia, N., \& Sramek, R.~A. 1990, \apj, 364, 611
\bibitem[Weiler, Panagia, \& Montes (2001)]{Weiler01} Weiler, K.\ W., Panagia, N., \& Montes, M. 2001, \apj, 562, 670
\bibitem[Weiler \etal (2002)]{Weiler02} Weiler, K.~W., Panagia, N., Montes, M.~J., \& Sramek, R.~A.\ 2002, \araa, 40, 387 
\bibitem[Weiler \etal (2007)]{Weiler07} Weiler, K.\ W., Williams, C.\ L., Panagia, N., Stockdale, C.\ J., Kelley, M.\ T., Sramek, R.\ A., Van Dyk, S.\ D., and Marcaide, J.\ M. 2007, \apj, submitted
\bibitem[Williams \etal (2002)]{Williams02} Williams, C.~L., Panagia, N., Van Dyk, S.~D., Lacey, C.~K., Weiler, K.~W., \& Sramek, R.~A.\ 2002, \apj, 581, 396 
\end{thebibliography}
\end{document}